\documentclass[twocolumn,aps,prb,notitlepage,showpacs,floatfix,unsortedaddress,superscriptaddress,citeautoscript]{revtex4-1}
\usepackage{amsmath,amssymb,amstext,graphicx,bm}
\usepackage{graphicx,color,bm}
\usepackage[english]{babel}
\usepackage{dsfont}
\usepackage{epsfig}
\begin{document}

\title{Revisiting the physical origin and nature of surface states in inverted-band semiconductors}

\author{Alexander Khaetskii}
\affiliation{Air Force Research Laboratory, Wright-Patterson AFB, Ohio 45433, USA}

\author{Vitaly Golovach}
\affiliation{Centro de F\'{i}sica de Materiales (CFM-MPC), Centro Mixto CSIC-UPV/EHU,  
20018 Donostia-San Sebasti\'{a}n, Spain}
\affiliation{Departamento de Pol\'{i}meros y Materiales Avanzados: F\'{i}sica, Qu\'{i}mica y Tecnolog\'{i}a, Facultad de Qu\'{i}mica, University of the Basque Country UPV/EHU, 
20080 Donostia-San Sebasti\'{a}n, Spain}
\affiliation{Donostia International Physics Center (DIPC),  
20018 Donostia-San Sebasti\'{a}n, Spain}
\affiliation{IKERBASQUE, Basque Foundation for Science, 48013 Bilbao, Spain}

\author{Arnold Kiefer}
\affiliation{Air Force Research Laboratory, Wright-Patterson AFB, Ohio 45433, USA}

\date{\today}

\begin{abstract}
We revisit the problem of surface states in semiconductors with inverted band structures, such as $\alpha$-Sn and HgTe. 
We unravel the confusion that arose over the past decade regarding the origin of the surface states, their topological nature, and the role of strain. Within a single minimalistic description, we reconcile different solutions found in the 1980s with the results obtained from modern-day numerical simulations, allowing us to unambiguously identify all branches of surface states around the $\Gamma$-point of the Brillouin zone in different regimes. We also show that strain is a smooth ``deformation'' to the surface states, following the usual \emph{continuity principle}  of physics, and not leading to any drastic change of the physical properties  in  these materials, in contrast to what has recently been advanced in the literature.  We consider biaxial in-plane strain that is either tensile or compressive, leading to different branches of surface states for topological insulators and Dirac semimetals, respectively. Our model can help in  interpreting  numerous experiments on topological surface states originating from inverted-band semiconductors. 
\end{abstract}

\maketitle

 \section{Introduction}      

Electronic surface states of solids have long been a topic of interest because of their importance in practical physical devices. The origin and properties of surface states for many semiconductor solids have been explained in great detail over several decades, beginning with Tamm and Shockley \cite{Tamm,Shockley}.  More recently, the prediction and discovery of topological insulators and their surface states have revived interest in this topic owing to their special properties \cite{Hasan,Qi}.  Present-day researchers often generate models of surface bands with sophisticated computational software, but they do not usually provide insight into the physical origin of these surface bands, often leading to a misguided treatment.  Furthermore, recent works seem unaware of the foundational analytical work conducted decades ago that provide similar physical predictions about surface states though at that time not recognizing them as having a specific topological character \cite{Dyak,Volkov,Cade,Suris,Volkov1}. 
\par
Topological insulators (TIs) serve as an excellent bridge between traditional electronic bandstructure theory and a more modern approach including topological aspects.  TIs emerge from a class of narrow-gap materials whose strong spin-orbit coupling leads to an inverted band structure and the formation of helical (i.e. spin-momentum locked) states on their surfaces \cite{Dyak,Volkov,Cade,Suris,Volkov1,Bernevig,Kane,Dai} that are protected from nonmagnetic perturbations. Growing attention to topological insulators is fueled by a fundamental interest in solid-state spin-physics and the prospect of designing novel electronic devices \cite{Ando}. Topological surface states have been observed in a number of 3D compounds, such as Bi$_2$Se$_3$, Bi$_2$Te$_3$, etc.~\cite{Zhang,Xia} Examples of 2D TIs with 1D helical channels at the sample edges include HgTe/CdHgTe \cite{Bernevig1,Konig} and InAs/GaSb \cite{Liu,Krishtopenko} quantum wells of certain thicknesses. Experimentally, surface states in 3D topological insulators are usually revealed by the angle- and spin-resolved photoemission spectroscopy \cite{Xia,Hsieh} and magneto-transport measurements \cite{Brune,Kozlov,Checkelsky,Analytis,Ren,Xia1,Pan}. 
\par
Strain is frequently present in real films and plays a role in revealing topological states by breaking crystalographic symmetries.  Several attempts to reveal experimentally the surface states in strained HgTe films (3D topological insulators) were made in magneto-transport experiments \cite{Brune,Kozlov}. It is interesting that the theoretical interpretation of the experimental data in these two works are based on very different physical pictures of the surface states in HgTe material. In particular, the surface states in Ref.~\onlinecite{Kozlov} are treated as linear in k-vector Dirac  states that exist in the gap between conduction s-band and light-hole p-band when the presence of the heavy-hole band is totally ignored. In other words, the topological surface states and bulk heavy-hole states are considered as completely independent.  In an opposing treatment offered in Ref.~\onlinecite{Brune}, strong hybridization between surface states and bulk heavy-hole states makes the physical picture totally different, where only the surface states lying within the strain-created gap between the light-hole and heavy-hole bands have physical meaning. Such a difference in interpretation requires further examination.  As we will explain in this work, we rather support the physical description of the surface states  used in Ref.~\onlinecite{Brune}. 
\par
\begin{figure}[!ht]
\vspace{0pt}\includegraphics[height=0.8\columnwidth]{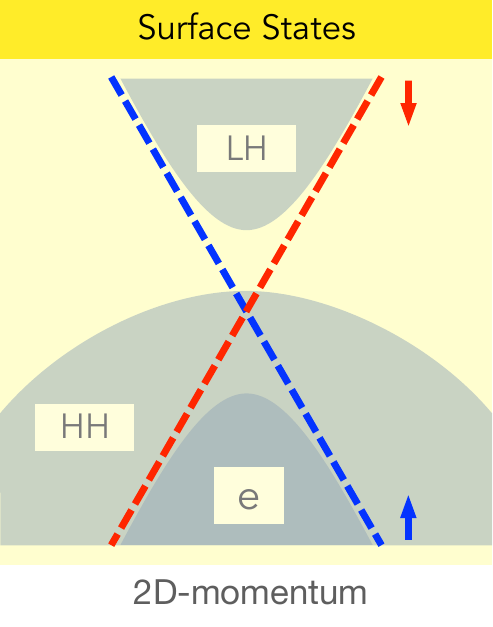}
\caption{\label{bands0}
Incorrect picture of surface states obtained after neglecting the coupling to the heavy-hole band (HH).
The surface states are formed in a two-band model consisting of electrons (e) and light holes (LH)
in an inverted-band semiconductor. 
}
\end{figure}

Another material attracting much attention now is $\alpha$-Sn, which has a band structure similar to HgTe.
Recently published research concerning surface states in $\alpha$-Sn (see, for example,  Refs.~\onlinecite{Ohtsubo,Barfuss,Kufner,Rogalev,Shia,Ding,Ding1}) clearly demonstrate the confusion that exists in the community about the role of band structure and strain. Many authors who treat the surface states by various numerical methods often ignore the presence of the heavy-hole band and solve the problem with only the light-hole and electron s-type bands. As we show below, this leads to an incorrect physical picture of the surface states, see Fig.~\ref{bands0}. Moreover, application of even a small  strain in their solutions causes a sudden appearence of the Dirac surface states deeply below the degeneracy point  ($\Gamma_8$) of the light and heavy holes, i.e. within the $\Gamma_8-\Gamma_6$  gap.  
This manifests the unphysical character of the obtained solutions. (We mean here a tensile in-plane strain applied to the inverted-band-gap semiconductor that opens a gap at the $\Gamma$-point between the light and heavy-hole bands).  
The troubles outlined above lead, in turn, to difficulties interpreting the experimental data obtained with the surface states in the materials like HgTe and $\alpha$-Sn.

\par
The goal of this work is to clarify the roles of band structure, strain, and spin in constructing a physical picture of surface states consistent with conventional solid-state theory. The authors of many results obtained solely by numerical calculation seem completely unaware of previous analytical works that treated surface states in similar materials (like HgTe) decades ago \cite{Dyak,Volkov,Cade,Suris,Volkov1}, making them difficult to relate.  We seek to reconcile these differences and show below that those older works contain all the essential physics and treat the band structure of the materials correctly  while still describing the surface states completely.  In particular,  we describe the intimate relation of the surface states in  HgTe gapless semiconductor first predicted within the Luttinger model in Ref.~\onlinecite{Dyak}  and surface states obtained within the more general Kane model that also takes the s-band into account. We demonstrate that these Dyakonov-Khaetskii (DK) states \cite{Dyak} possess all the essential features of the topological states  in the presence of strain which induces a gap between the light and heavy-hole bands (topological insulator regime). As such, they should play the most  important role in the experiments that probe the topological properties of the sample when the Fermi level is located within this strain-induced gap. Moreover, we show that strain does not drastically change the properties of surface states that form the Dirac cone within the $\Gamma_8-\Gamma_6$  gap, which is again in strong contrast to the results advanced recently in the literature. 
\par
\begin{figure}[!ht]
\vspace{0pt}\includegraphics[height=0.8\columnwidth]{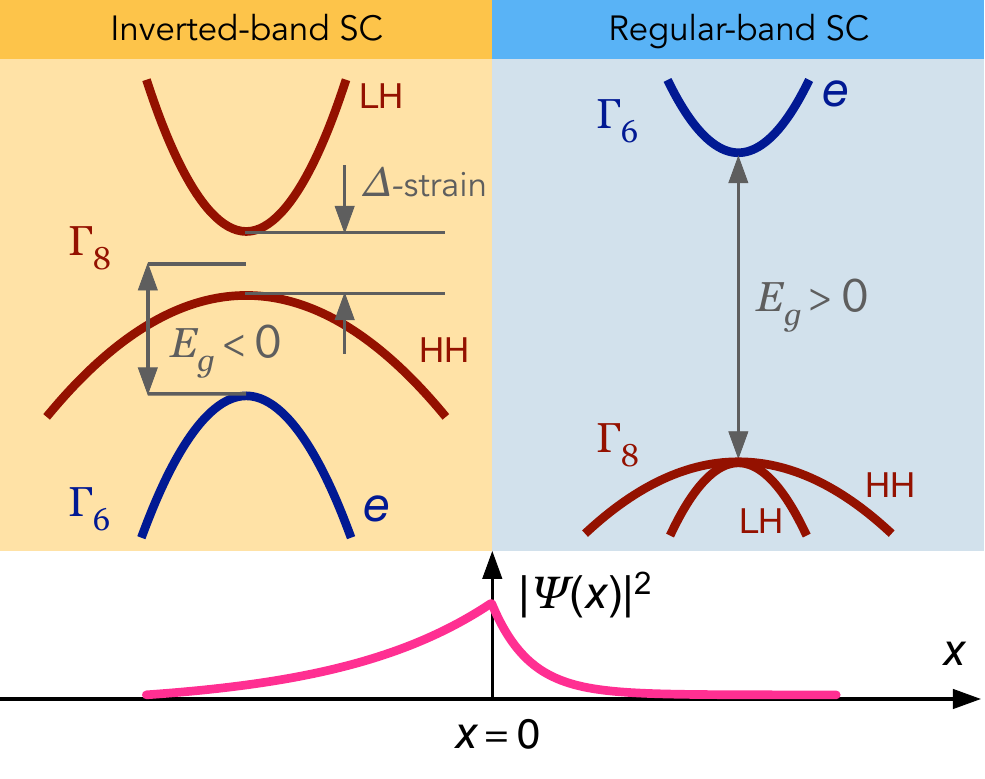}
\caption{\label{bands}
Generic band structure variation across an interface between an inverted-band-gap semiconductor (SC) (left: $E_g<0$) and a SC with a regular band order (right: $E_g>0$). A tensile in-plane strain applied to the inverted-band-gap SC induces a splitting between the light-hole and heavy-hole states and opens a gap $\Delta$ at the $\Gamma$-point.  
At certain energies (not shown), 
the interface supports surface states with a probability density which exponentially decays into both regions, thus being localized at the interface (bottom panel). 
}
\end{figure}

 We take a pedagogical approach, beginning with a simple, idealized model that captures the essential physics. 
As such we consider a variant of the Kane k-P model with infinite mass of heavy holes in the case of a single abrupt interface between an inverted  material like HgTe and a direct material like CdTe or vacuum. This model takes into account three bands: an s-type conduction band and p-type light and heavy hole bands, see Fig.~\ref{bands}. The split-off band is relatively far away in energy (strong spin-orbit coupling) and does not play a significant role in forming the surface states.  By considering the flat heavy-hole band  we deal with the case of a real gap between hole bands ($\Gamma_8$) and electron s-band ($\Gamma_6$) when the density of states within this gap is zero.  These features of the model make the physics of the surface states located in the $\Gamma_8-\Gamma_6$  gap very simple and reveals their essential features. We then address the realistic case of finite heavy-hole mass and show the surface states located within the $\Gamma_8-\Gamma_6$  gap overlap the bulk heavy-hole states in energy and are strongly hybridized with them, making those surface states marginal in observable phenomena.  
\par
For a detailed description of the surface states that exist within the projected gap between the light and heavy-hole bands (DK states) we use the Luttinger model which is applicable for the case of arbitrary masses of light and heavy holes.
 We were able to find exact analytical solutions for the surface states within this model for both signs of the strain energy $\Delta$, covering the TI regime ($\Delta>0$) and the Dirac semimetal regime ($\Delta<0$). In the latter case, which is achieved by applying a compressive in-plane strain (hence tensile in the normal direction), an intersection of the light-hole and heavy-hole bands occurs forming two Dirac points with linear bulk dispersion nearby. This Dirac semimetal phase has been attracting a lot of attention recently both theoretically and experimentally, see  Refs.~(\onlinecite{Ding,Ding1}).
We show that the dispersion of surface states  which arise around the conic point of the bulk projected states crucially depends on the applied strain, in contrast to the TI regime. 
\par
 Our simple models  allow us to trace the origins of the incorrect physical picture of the surface states obtained in many recent publications, namely, neglecting the coupling to the heavy holes band that leads directly to the simple picture of linear in k-vector  Dirac surface states crossing the whole gap between the conduction and  light-hole bands, see Fig.~\ref{bands0}. 
 We stress again that this latter picture does not reflect the true nature of the surface states. 

 \section{The $3\times 3$ Kane model, an abrupt interface of two materials -with inverted and direct bands}    
 \label{3x3}

We consider here the simplest possible problem: the interface between two materials, one of them having an inverted band structure and the other -a direct one.  Our description is quite general and is applicable for many materials, but as a concrete example one can consider the (HgTe/CdTe) case. 
Additionally, we assume a biaxial tensile in-plane strain applied to the inverted band material as is often accomplished in practice by epitaxial growth of a thin film on a much thicker substrate, see Fig.~\ref{bands}.
 The band structures are described by the $6\times 6$ Kane model that takes into account three bands, electrons (e) (s-symmetry band), and two p-bands: light  (lh) and heavy holes (hh).  To simplify the treatment, we can from the very beginning use the time-reversal character of the problem and reduce the model to $3\times 3$ by choosing the proper coordinate system \cite{Subashiev}. 
We choose the angular momentum quantization axis z in the plane of the interface and the direction of the carrier motion along y; x denotes the direction normal to the interface. Then the two groups of states  (e 1/2, lh -1/2, hh 3/2) and (e -1/2, lh 1/2, hh -3/2) do not mix. The Hamiltonian matrix for the first group of states is
\begin{eqnarray}
\hat{H}&=&
\begin{bmatrix}
\epsilon_{c1,2} & Pk_-/\sqrt{6} & Pk_+/\sqrt{2} \\
Pk_+/\sqrt{6}  & \epsilon_{v1,2} -\Delta/4 &
\sqrt{3}\Delta/4\\
Pk_-/\sqrt{2} &\sqrt{3}\Delta/4 &\epsilon_{v1,2} +\Delta/4
\end{bmatrix}
\label{Hamiltonian}
\end{eqnarray}
(The Hamiltonian matrix that refers to the second group of states with the opposite sign of the angular momentum projection on the z-axis can be obtained from Eq.~(\ref{Hamiltonian}) by replacing $k_y$ by $-k_y$). 
In Eq.~(\ref{Hamiltonian}) $P$ is the Kane matrix element which we consider coordinate independent, $\hat{k}_{\pm}=\hat{k}_x \pm ik_y$, $\hat{k}_x=-id/dx$, $\epsilon_{c1,2}, \epsilon_{v1,2}$ are conduction- and valence-band center energies in inverted-band semiconductor I and in semiconductor II with direct bands, respectively. 
We assume that HgTe layer is strained due to the lattice mismatch between CdTe and HgTe and this tensile in-plane strain opens the gap  $\Delta >0$ between the light and heavy-hole bands. If the CdTe is oriented along (001), then the strain tensor components are given by $\epsilon_{yy}=\epsilon_{zz}=(a_{\textrm{CdTe}}-a_{\textrm{\textrm{HgTe}}})/a_{\textrm{HgTe}}>0 $ (tensile in-plane strain), 
$\epsilon_{xx} <0$; the non-diagonal components are equal to zero. Here $a_{\textrm{CdTe}}$ and $ a_{\textrm{HgTe}}$ are the lattice constants of the corresponding materials. 
The effects of the strain tensor are incorporated in the Kane model through the Bir-Pikus Hamiltonian \cite{Bir}, which is easily obtained from the Kane Hamiltonian with the substitution $k_ik_j \rightarrow \epsilon_{ij}$. The value of the strain gap in our case is given by $\Delta=2b(\epsilon_{xx}-\epsilon_{yy}) >0$, where $b$ is the uniaxial deformation potential ($b=-1.5 eV$ for HgTe).
\par
With  the proper unitary transformation one can bring the Hamiltonian Eq.~(\ref{Hamiltonian}) to the following form:
 \begin{eqnarray}
\hat{H}&=&
\begin{bmatrix}
\epsilon_{c1,2} & \frac{ (2\hat{k}_x+ik_y)P}{\sqrt{6}} & iPk_y/\sqrt{2} \\
 \frac{(2\hat{k}_x-ik_y)P}{\sqrt{6}}  & \epsilon_{v1,2} +\Delta/2&
0\\
-iPk_y/\sqrt{2} & 0 &\epsilon_{v1,2} -\Delta/2
\end{bmatrix}
\label{Hamilt}
\end{eqnarray}

\subsection{No coupling to the heavy-hole band}
\label{no_coupl}

Before proceeding further we can make here the following important note which explains the origin of the wrong physical picture of the surface states obtained in Refs.~\onlinecite{Ohtsubo,Barfuss,Kufner,Rogalev}. If one neglects altogether the interaction of the $\Gamma_6$ band with the heavy-hole band, given by elements $H_{13}$ and $H_{31}$  in Eq.~(\ref{Hamilt}), then one obtains the anisotropic Dirac model which describes the interaction of the two bands, electrons and light holes, as described by the 2x2 block in the upper left corner of the matrix in Eq.~(\ref{Hamilt}).  Let us find the energy dispersion of the surface states for this reduced case, considering semi-infinite sample of HgTe located at $x<0$ and CdTe sample located at $x>0$. The wave functions decay in the left-hand region  as $\exp(\kappa_1 x)$ and in the right-hand region as $\exp(-\kappa_2 x)$ (i.e. $k_{x1}=-i\kappa_1, k_{x2}=i\kappa_2$, with positive $\kappa_{1,2}$); the dependence along the y-coordinate is given by $\exp(ik_yy)$. For simplicity we consider here the symmetric case $\Delta=0, \epsilon_{v1}-\epsilon_{c1}=\epsilon_{c2}-\epsilon_{v2}=\epsilon_g>0$. Then from the bulk energy dispersion 
$\epsilon_b^2=\epsilon_g^2/4+(P^2/6)(k_y^2+4k_x^2)$
we easily find $\kappa_1=\kappa_2=\kappa=(1/2)\sqrt{k_y^2+ (6/P^2) (\epsilon_g^2/4-\epsilon^2)}$, where $\epsilon$ is the energy of the surface state obtained  from the boundary conditions. As the boundary conditions we use the continuity of  both components of the two-component spinors found in regions I and II. For our symmetric case it gives the following equation:   
$-4\kappa \epsilon=k_y \epsilon_g $. Using the expression for the $\kappa$, we finally obtain the equation for the surface states energy
\begin{equation}
\epsilon \sqrt{k_y^2+ \frac{6}{P^2} \big(\frac{\epsilon_g^2}{4}-\epsilon^2\big)}=-\frac{k_y}{2}\epsilon_g
\label{Dirac}
\end{equation}
This equation has the exact solution $\epsilon=-Pk_y/\sqrt{6}$. (The second branch of surface states is obtained by changing the sign of $k_y$). Note that the surface curve  $\epsilon=Pk_y/\sqrt{6}$ asymptotically approches the bulk light-hole dispersion curve at large $k_y$ and $k_x=0$. Thus we see that the Dirac solution for the surface states is obtained by total neglect of the interaction with the heavy-hole band.  

\subsection{Full treatment}
Let us now find the correct energy spectrum of the surface states for the same problem with heavy-hole interactions included.
As it follows from Hamiltonian (\ref{Hamilt}), 
an interaction of the $\Gamma_6$ band with the heavy-hole band is given by the $ik_yP/\sqrt{2}$ element, which does not depend on $d/dx$. For that reason we can exclude the heavy hole component of the wave function and still work with the two-band model, i.e. electrons and light  holes.  It is easy to work out from Eq.~(\ref{Hamilt}) that this procedure leads to the system of two equations in each region: $\hat{H}_{eff}\,\, \chi=0$, $\chi_{tr}=(u,v)$ with the ``effective'' Hamiltonian (see also Ref.~\onlinecite{Volkov1}):
\begin{eqnarray}
\hat{H}_{eff}&=&
\begin{bmatrix}
\epsilon_{c1,2}- \epsilon -\frac{P^2k_y^2 }{2(\epsilon_{v1,2}-\epsilon-\Delta/2)}& \frac{ (2\hat{k}_x+ik_y)P}{\sqrt{6}} \\
 \frac{(2\hat{k}_x-ik_y)P}{\sqrt{6}}  & \epsilon_{v1,2}-\epsilon  +\Delta/2 
\end{bmatrix}
\label{effective}
\end{eqnarray}
We assume that strain is zero in region II  and $\Delta$ is positive in region I, i.e. a gap is open between the light and heavy-hole bands (see Fig.~\ref{bands}).   
From Eq.~(\ref{effective}) we find the equation for the energy dispersions of the bulk materials:
\begin{eqnarray}
[\epsilon_{v1,2}-\epsilon -\frac{\Delta}{2}]\cdot [(\epsilon_{c1,2}-\epsilon)(\epsilon_{v1,2}-\epsilon +\frac{\Delta}{2}) 
\nonumber \\ 
-\frac{2}{3}P^2(k_x^2+k_y^2)]=P^2k_y^2 \frac{\Delta}{2}
\label{bulk}
\end{eqnarray}
We see that in the absence of strain the spectrum of heavy holes with infinite mass is simple: $\epsilon_h=\epsilon_{v1,2}$.
Strain mixes the light and heavy holes, the heavy holes acquire finite mass in the y-direction (within the plane), and the spectrum becomes complicated. In the x-direction, however, when $k_y=0$, the spectra of light and heavy particles remain simple:
\begin{eqnarray}
 (\epsilon_{c1,2}-\epsilon(k_x))(\epsilon_{v1,2}-\epsilon(k_x) +\frac{\Delta}{2})&=&\frac{2}{3}P^2k_x^2
\label{bulkl} \\ 
\epsilon_{h1,2}(k_x)&=&\epsilon_{v1,2}-\frac{\Delta}{2}
\label{bulkh}
\end{eqnarray}
Equation~(\ref{bulkl}) gives the bulk spectrum of the light  particles---electrons and light holes, 
with dispersions $\epsilon_e(k_x)$ and $\epsilon_l(k_x)$, respectively.
\par
We seek for the surface state that is localized at the interface ($x=0$ ) and decays in the left-hand  region  as $\exp(\kappa_1 x)$ and in the right-hand region as $\exp(-\kappa_2 x)$ ( $\kappa_{1,2}>0$). The corresponding expressions for the $\kappa_{1,2}$  can be found from Eq.~(\ref{bulk}):
 \begin{eqnarray}
\kappa_1^2&=&k_y^2+ \frac{3(\epsilon- \epsilon_{c1})(\epsilon_{v1}-\epsilon +\frac{\Delta}{2})}{2P^2}+\frac{3\Delta k_y^2}{4(\epsilon_{v1}-\epsilon -\frac{\Delta}{2})}; \label{k1}  \\
\kappa_2^2&=&k_y^2+ \frac{3(\epsilon_{c2}- \epsilon)(\epsilon-\epsilon_{v2})}{2P^2}.
\label{k2}
\end{eqnarray}
The spinors in both regions (I) and (II) can be easily found with the help of Eq.~(\ref{effective}); then from the continuity  conditions of both components of the wave functions  at $x=0$ we find:
\begin{equation}
(\epsilon_{v1}-\epsilon +\frac{\Delta}{2})(2\kappa_2-k_y)=(\epsilon-\epsilon_{v2})(2\kappa_1+k_y).
\label{BCs}
\end{equation} 
The surface states spectra are determined by the solution of the system of equations (\ref{k1},\ref{k2},\ref{BCs}).

\begin{figure*}[!ht]
\vspace{0pt}\includegraphics[height=20em]{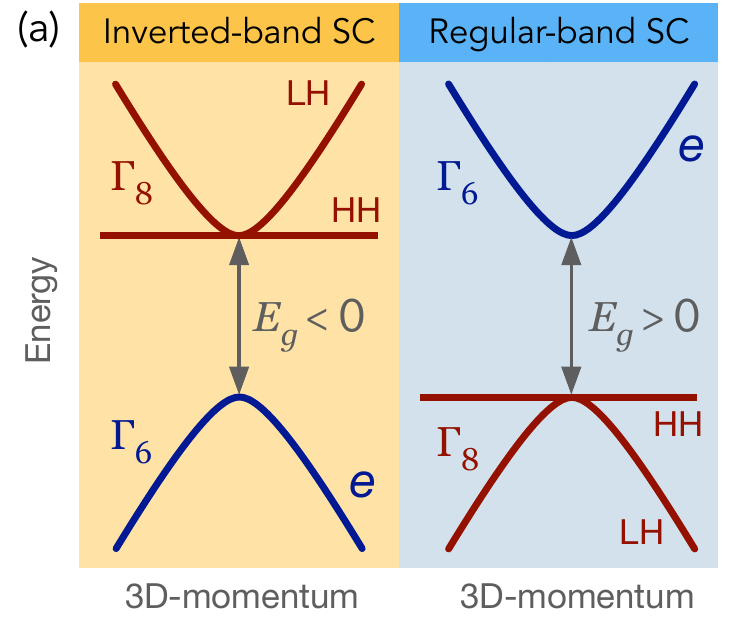}
\hspace{12pt}
\vspace{0pt}\includegraphics[height=20em]{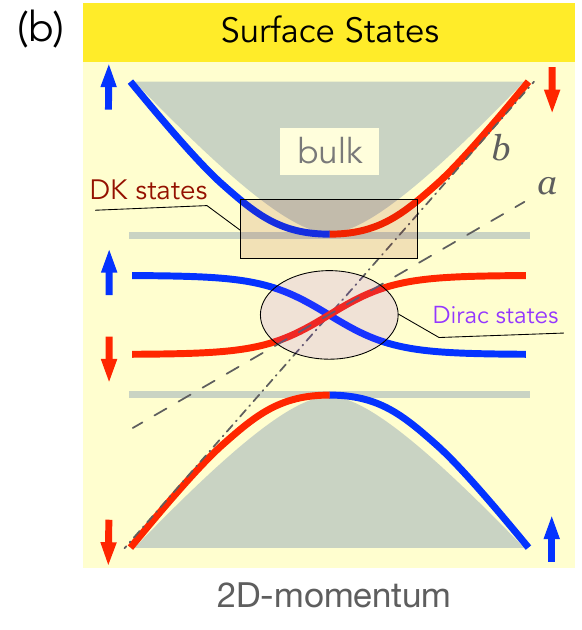}
\caption{\label{symmetricCase}
 (a) An interface with an antisymmetric variation of the bulk band structure. 
 For simplicity, the strain is omitted here and the heavy-hole mass is set to infinity. 
 (b) The surface states arising from the band alignment in (a), showing
 two types of states near the center of the Brillouin zone: 
 linear Dirac states and quadratic Dyakonov-Khaetskii states (DK states).
 The projected bulk spectrum is shown by the shaded gray regions.
 All surface states are singly-degenerate
 with a distinct spinor structure originating from one of the two time-reversal-related blocks of the Hamiltonian
 as shown by the spin-up and spin-down arrows.
 The asymptote of the Dirac states at small in-plane momenta (dashed line $a$)
 differs strongly from the asymptote of the DK states at large momenta (dot-and-dashed line $b$), 
 refuting the picture of a narrow-in-energy interruption of the  Dirac states by the heavy-hole band.
 }
\end{figure*}

\subsubsection{Symmetric case, $\Delta=0$, $|\epsilon_{g1}|=\epsilon_{g2}=\epsilon_g $}
The discussion of the solution for the surface states within the full treatment  starts with the symmetric case (Fig.~\ref{symmetricCase}(a))
 $\Delta=0$, $\epsilon_{v1}=-\epsilon_{c1}=\epsilon_{c2}=-\epsilon_{v2}=\epsilon_g/2>0$ so we may compare it with the results obtained in section \ref{no_coupl}.  In this case we have $\kappa_1=\kappa_2=\kappa=\sqrt{k_y^2+ (3/2P^2) (\epsilon_g^2/4-\epsilon^2)}$, and from Eq.~(\ref{BCs}) one has the same equation as before,  $-4\kappa \epsilon=k_y \epsilon_g $, therefore finally we obtain
\begin{equation}
\epsilon \sqrt{4 k_y^2+ \frac{6}{P^2} \big(\frac{\epsilon_g^2}{4}-\epsilon^2\big)}=-\frac{k_y}{2}\epsilon_g.
\label{symm}  
\end{equation}
(The time-reversed solutions can be  obtained by changing the sign of $k_y$ in Eq.~(\ref{symm})). 
Despite the fact that Eq.~(\ref{Dirac}) and Eq.~(\ref{symm})
are nearly the same (except for the coefficient of the first term under the radical), they have completely different solutions, see Fig.~\ref{symmetricCase}(b).
For $|k_y| \ll \epsilon_g/P$ Eq.~(\ref{symm}) has the same behavior: $\epsilon=-Pk_y/\sqrt{6}$; however, at large $|k_y| \gg\epsilon_g/P $ the surface branch saturates at $ +\epsilon_g/4$ (at negative $k_y$) and at $ -\epsilon_g/4$ (at positive  $k_y$). Similar behavior was found in Refs.~\onlinecite{Cade,Suris,Volkov1}. This behavior is due to an interaction with the heavy-hole band which has the character of repulsion.

\par
Besides the surface states that
 lie within the fundamental gap, Eq. (\ref{symm}) has the solutions which start from the heavy-hole bands $(\epsilon=\epsilon_g/2, \epsilon=-\epsilon_g/2) $, see Fig.~\ref{symmetricCase}(b). For the first of those additional solutions, for example, we obtain at small $|k_y|P\ll \epsilon_g$
\begin{equation}
\epsilon_{s1}=\frac{\epsilon_g}{2}+ \frac{P^2k_y^2}{2\epsilon_g}.
\label{DKh}
\end{equation}
This solution follows  from Eq. (\ref{symm}) at $k_y<0$. (For $k_y>0$ the solution with this energy does not exist since the left-hand side of Eq.~(\ref{symm}) is strictly positive).  The same energy dispersion for $k_y>0$ follows from the time-reversed Hamiltonian and corresponds to the opposite spin direction. It means that there is a momentum-spin locking for these solutions, and back scattering is strictly forbidden.    We will discuss this issue in more detail below. 
The effective mass of this surface state, Eq.~(\ref{DKh}),  is by the factor of $4/3$  larger than the effective mass of the light hole $m_l$. (The latter is found from the bulk dispersion equation for the light holes: $\epsilon_l(k_y)=\sqrt{\epsilon_g^2/4+ (2/3)P^2 k_y^2} $ ). This new surface branch was predicted theoretically  for HgTe for the first time in Ref.~\onlinecite{Dyak}. The effective mass of this surface state depends on the ratio $\beta$ of the bulk effective  masses of light and heavy holes and approaches the value $(4/3) m_l$ (see Ref.~\onlinecite{Dyak}) in the limit $\beta \to 0$ considered  here. As it follows from Eq.~(\ref{symm}) (and from the time-reversed one), at large $|k_y| \gg \epsilon_g/P$ the surface branch in question asymptotically approaches the bulk light-hole band: $\epsilon_{s1}\to \sqrt{2/3}P|k_y| + (3\sqrt{3}/32\sqrt{2})(\epsilon_g^2/P|k_y|)$.

 \subsubsection{Evolution of surface states with the strength of interaction between $\Gamma_6$ and heavy-hole band}

\begin{figure}[!ht]
\vspace{0pt}\includegraphics[height=14em]{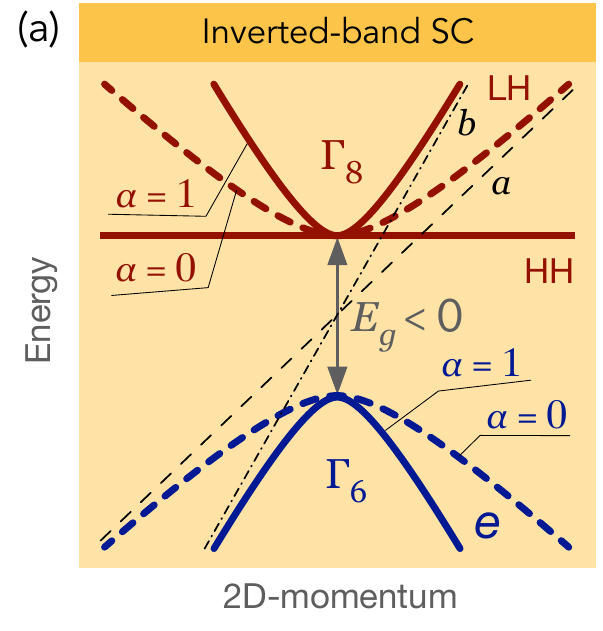}
\vspace{0pt}\includegraphics[height=14em]{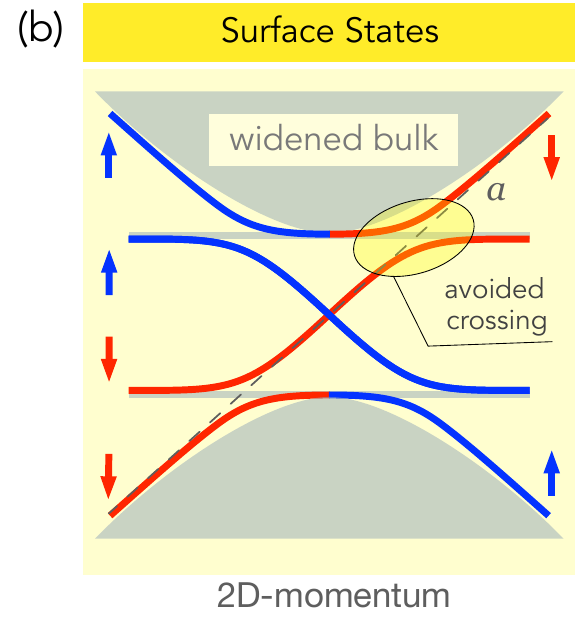}
\caption{\label{evolutionCase}
Dirac cone influenced by a weak coupling to the heavy-hole band.
(a) The widening of the light-hole (LH) and electron (e) branches in the in-plane momentum direction
upon weakening of the coupling to the heavy-hole branch.
The coupling to the heavy-hole branch is full for $\alpha=1$ (solid lines), corresponding to a pristine semiconductor, 
and absent for $\alpha=0$ (dashed lines), when dispensing with the heavy-hole branch.
The asymptotes $a$ and $b$ delimit the bands for $\alpha=0$ and $1$, respectively. 
(b) The surface states obtained in the limit $\alpha\ll 1$, showing an avoided crossing of the Dirac cone with the heavy-hole bands. 
Here, we consider the ``symmetric case'' of Fig.~\ref{symmetricCase} for a small $\alpha$ introduced alike for the 
inverted-band-gap and regular semiconductors. 
}
\end{figure}

To  better understand the  origin of different branches of surface states presented in Fig.~\ref{symmetricCase}(b), it is instructive to follow the evolution of these states as a function of the strength of the interaction of the $\Gamma_6$ band with the heavy-hole band.  To this end we multiply the corresponding matrix elements in Eq.~(\ref{Hamilt}) by the factor $\alpha$, where $\alpha$ can take the values in the interval between 0 and 1, obtaining the Hamiltonian 
 \begin{eqnarray}
\hat{H}_{\alpha}&=&
\begin{bmatrix}
\epsilon_{c1,2} & \frac{ (2\hat{k}_x+ik_y)P}{\sqrt{6}} & i\alpha Pk_y/\sqrt{2} \\
 \frac{(2\hat{k}_x-ik_y)P}{\sqrt{6}}  & \epsilon_{v1,2} +\Delta/2&
0\\
-i\alpha Pk_y/\sqrt{2} & 0 &\epsilon_{v1,2} -\Delta/2
\end{bmatrix}
\label{Hamilt_alpha}
\end{eqnarray}
Then for the symmetric case we obtain the following equation which describes the surface energy spectrum
\begin{equation}
\epsilon \sqrt{(1+3\alpha^2) k_y^2+ \frac{6}{P^2} \big(\frac{\epsilon_g^2}{4}-\epsilon^2\big)}=-\frac{k_y}{2}\epsilon_g.
\label{alpha}  
\end{equation}
This equation reproduces Eq.~(\ref{Dirac}) and Eq.~(\ref{symm}) in the limits $\alpha\to 0$ and $\alpha \to 1$, correspondingly. 
It is useful to take the square of this equation to obtain 
\begin{equation}
(\frac{\epsilon_g^2}{4}-\epsilon^2 ) (k_y^2- \frac{6\epsilon^2}{P^2})=3\alpha^2k_y^2\epsilon^2
\label{alpha1}
\end{equation}
Let us investigate the solutions in the $\alpha \to 0$ limit. There are two of them with the anticrossing point 
$\tilde{k}_y=\sqrt{3/2}(\epsilon_g/P)$, see Fig.~\ref{evolutionCase}(b).  The first solution has the dispersion $\epsilon=\epsilon_g/2 + (\alpha^2P^2k_y^2/2\epsilon_g)$ at small $k_y$. (Note that this dispersion coincides with the one given by Eq.~(\ref{DKh})) at $\alpha=1$). At large $k_y$ this branch is close to the Dirac solution $Pk_y/\sqrt{6}$. The second solution coincides with the Dirac   one at small $k_y$ and is close to the heavy-hole band ($\epsilon_g/2$) at large $k_y$. These two branches repel each other with increasing $\alpha$ and eventually approach the configuration shown in  Fig.~\ref{symmetricCase}(b) (red curves).
 For example, the asymptote of the DK state at large $k_y$ is $Pk_y \sqrt{(1+3\alpha^2)/6}|_{\alpha=1}=Pk_y\sqrt{2/3}$ (asymptote b in Fig.~\ref{symmetricCase}(b)). 
\par
Thus we see that the surface states presented in Fig.~\ref{symmetricCase} (see also Fig.~\ref{highBarrierCase} below) are intimately related to the  Dirac states. 
 In particular,  DK surface state  \cite{Dyak} predicted for the first time within the Luttinger model  is
 the result of strong hybridization of the  Dirac state with the heavy-hole states (see also Ref.~(\onlinecite{Krishtopenko1})). 
The  authors of Ref.~\onlinecite{Brune} claimed the same, though they referred to the states only as ``special'' rather than identify them as the  DK states previously established in Ref.~\onlinecite{Dyak}.

\subsubsection{High barrier between inverted and direct materials}

\begin{figure}[!ht]
\vspace{0pt}\includegraphics[height=14em]{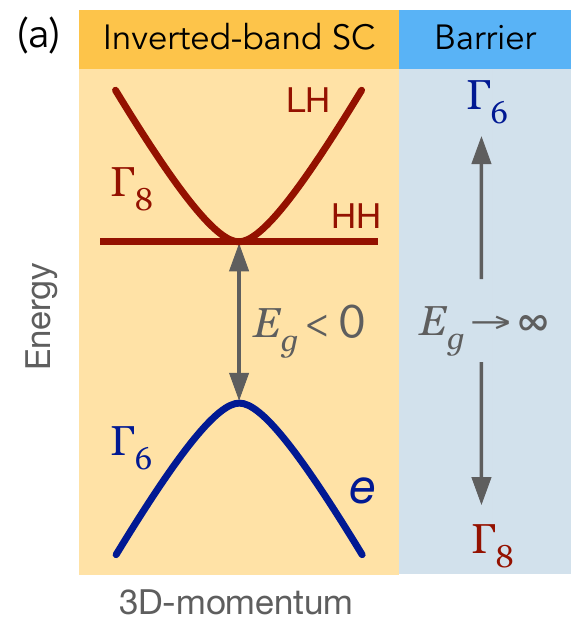}
\vspace{0pt}\includegraphics[height=14em]{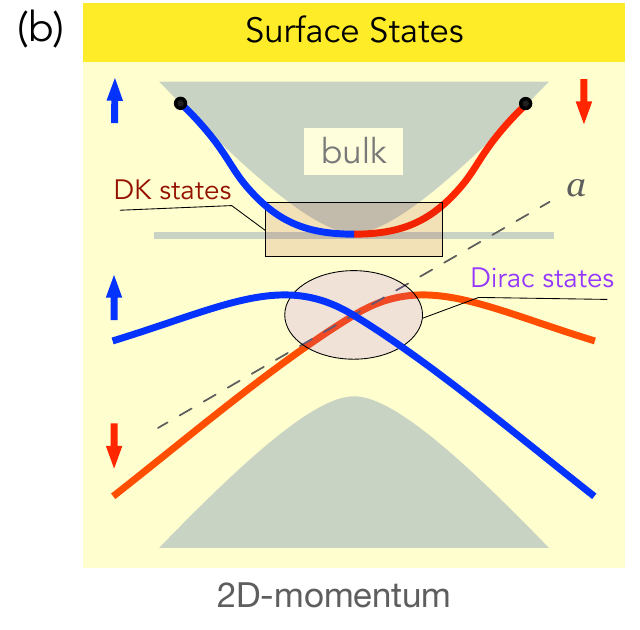}
\caption{\label{highBarrierCase}
 Effect of a high barrier on the surface states.
 (a) An interface between an inverted-band-gap SC ($E_g<0$) and a regular SC with a large fundamental gap ($E_g\to +\infty$).
 (b) Similar to Fig.~\ref{symmetricCase}, two types of surface states occur in the limit of small in-plane momenta,
 namely the linear Dirac states and the quadratic DK states.
 At large momenta, the lower cone of the Dirac states becomes sharper than the original low-momentum asymptote $a$,
 whereas the upper cone reverses slope and becomes blunter.
The DK states terminate at a critical value of momentum $k_y\sim \epsilon_g/P$ (see solid circles), at which the localization length diverges ($\kappa_1\to 0$) and the surface state \emph{flows} tangentially into the bulk continuum.      
}
\end{figure}

The surface states spectra depend strongly on the height of the barrier between inverted and direct materials. If the fundamental gap in the direct material $\epsilon_{g2}$ is much larger than the modulus of the gap in the inverted one, then 
 Eq.~(\ref{BCs}) (and time-reversed one) read
\begin{equation}
\frac{\epsilon_{g1}}{2}-\epsilon -\sqrt{(\frac{\epsilon_{g1}}{2})^2 -\epsilon^2 +\frac{2}{3}P^2 k_y^2}=\pm \frac{Pk_y }{\sqrt{6}}.
\label{high}
\end{equation}
As before, we first consider here $\Delta=0$ case and symmetric energy diagram: $\epsilon_{v1}=-\epsilon_{c1}=\epsilon_{g1}/2 >0;
\epsilon_{c2}=-\epsilon_{v2}=\epsilon_{g2}/2 \to \infty$. Equation (\ref{high})  can be easily solved analytically  and surface curves are presented in  Fig.~\ref{highBarrierCase}. We see that the surface branches which pass through the center of the gap (Dirac point) do not saturate at large $|k_y|$ but go downward. It happens because the heavy-hole band in the barrier region is located deeply below in energy which strongly suppresses  repulsion from it.  
These two surface branches have the following exact dispersions
\begin{eqnarray}
\epsilon_{\downarrow}=\frac{\epsilon_{g1}}{4}+\frac{Pk_y}{2\sqrt{6}}-\sqrt{(\frac{\epsilon_{g1}}{4})^2 -\frac{Pk_y\epsilon_{g1}}{4\sqrt{6}} +\frac{7}{24}P^2 k_y^2}  \nonumber \\
\epsilon_{\uparrow}=\frac{\epsilon_{g1}}{4}-\frac{Pk_y}{2\sqrt{6}}-\sqrt{(\frac{\epsilon_{g1}}{4})^2 +\frac{Pk_y\epsilon_{g1}}{4\sqrt{6}} +\frac{7}{24}P^2 k_y^2} \nonumber \\
\label{Dirac1}
\end{eqnarray}
As for the DK surface branch, it has the  quadratic dispersion for  small $k_y P\ll \epsilon_{g1}$ with the same effective mass as it was for the case of equal gaps of direct and inverted materials, see Eq.~(\ref{DKh}). The behavior of this branch at larger $k_y$ can be traced with the exact solution of Eq. (\ref{high}) (for positive $k_y$ we take the minus sign in the right-hand side of this equation). This solution reads
\begin{equation}
\epsilon_{s1}=\frac{\epsilon_{g1}}{4}+\frac{Pk_y}{2\sqrt{6}}+\sqrt{\left (\frac{\epsilon_{g1}}{4}\right )^2 -\frac{Pk_y\epsilon_{g1}}{4\sqrt{6}} +\frac{7}{24}P^2 k_y^2}; \,\, k_y\geq 0.
\label{high1}
\end{equation}
It can be easily checked from Eq. (\ref{high}) that this solution is only valid at $Pk_y\leq \sqrt{2/3}\epsilon_{g1}$. 
At this critical $k_y$ the DK surface state merges with  the light-hole band at the energy value  which is equal to $(5/6)\epsilon_{g1}$, see Fig.~\ref{highBarrierCase}(b).

\subsubsection{Finite effective mass of the heavy-hole band}

So far we have been considering the case when an effective mass of the heavy-hole (HH) band is infinite.  
It is important to realize that  in this limit there is a true gap (i.e. density of states is zero)  between the valence $\Gamma_8$ bands and conduction $\Gamma_6$ band, Figs.~(\ref{symmetricCase},\ref{highBarrierCase}). When effective mass of heavy holes is finite, their energy starts to depend on the $k_x$-wave vector component perpendicular to the  interface. It means that the surface states located within the $\Gamma_8-\Gamma_6$ gap  lie on top of the continuum of the heavy-hole states corresponding to different $k_x$ and strong hybridization  between them and surface states will occur. As a result, the electrons occupying the surface states located within the $\Gamma_8-\Gamma_6$ gap will not show in transport measurements the properties exclusive for topological surface states not hybridized with any bulk states. Thus, they cannot be distingushed  as a separate group of carriers. (We should mention, however, that at small $k_y$, at which the overlap of these states with heavy holes is weak, they can be detected with the ARPES experiments, see Refs.~\onlinecite{Rogalev,Rauch1}). 
 On the other hand, the surface states lying in the projected gap between the light-hole and heavy-hole bands found in Ref.~\onlinecite{Dyak} have  true physical meaning. (The similar ideas about the hybridization and its effect on the surface states were developed in Ref.~\onlinecite{Brune}). Now we will study how these states are transformed in the presence of the strain which opens a true gap between the light-hole and heavy-hole bands. 

\subsection{Strain applied to the inverted region}
We now analyze the same surface-states problem with strain applied to the inverted region, thereby opening the gap $\Delta$ between light-hole and heavy-hole bands,  see Fig.~\ref{bands}. We consider the case $\Delta \ll \epsilon_{g1}$, which is of practical importance, and will assume first the symmetric case, $\epsilon_{v1}=-\epsilon_{c1}=\epsilon_{c2}=-\epsilon_{v2}=\epsilon_g/2>0$. 
Clearly, the physical picture of the states lying within a  $\Gamma_8-\Gamma_6$ gap cannot change appreciably with application of strain, see Figs.~(\ref{symmetricCase},\ref{highBarrierCase}). Indeed, the perturbation theory with respect to the parameter $\Delta/\epsilon_g \ll 1$ shows that the dispersion curves within the  $\Gamma_8-\Gamma_6$ gap shift only slightly (for example, the Dirac point shifts from the middle of the gap by the value of the  order of $\Delta$).   
The only strong change of the spectrum occurs in the energy interval of the order of $\Delta$ near the top of the valence band of HgTe. Strain opens a gap; therefore, when the Fermi level is located within this gap, there is no conductivity (at low temperature) in the bulk and no elastic scattering between surface and bulk states. We stress, however, that the topological character of this surface state, Eq.~(\ref{DKh}),  exists even in the absence  of the strain (as we have already mentioned above). In particular, all the characteristic features of the topological state, such as a spin-momentum locking and absence of the back scattering, exist already for the  solution, Eq.~(\ref{DKh}), obtained  without strain.
\par
Let us find the energy spectrum of the DK surface state in the presence of  strain. 
From the system of equations (\ref{k1},\ref{k2},\ref{BCs}) it is easy to find that at $k_y \to 0$ the solution has the form $\epsilon=\epsilon_{v1}-\Delta/2+ \delta P^2k_y^2$ with $\kappa_1^2=(3\Delta/2P^2)(\epsilon_g-\Delta/2-1/(2\delta))$ and 
$\kappa_2^2=(3\Delta/4P^2)(\epsilon_g-\Delta/2)$. Then from Eq. (\ref{BCs}) we can calculate $\delta$ and, finally, the surface state spectrum at $k_y \to 0$ reads
\begin{equation}
\epsilon=\epsilon_{v1}-\frac{\Delta}{2}+\frac{ P^2k_y^2}{2\epsilon_g-\Delta -\frac{2\Delta^2}{(2\epsilon_g-\Delta)}}
\label{strain}
\end{equation}
We note that Eq.~(\ref{strain}) is valid at arbitrary values of strain gap $\Delta$, not necessarily much smaller than $\epsilon_g$. 
\par
For the case of a very large gap of the direct material ($\epsilon_{g2}\to \infty$), we obtain a similar equation for the surface state dispersion at $k_y \to 0$ 
 \begin{equation}
\epsilon=\epsilon_{v1}-\frac{\Delta}{2}+\frac{ P^2k_y^2}{2\epsilon_{g1}-3\Delta} 
\label{strain1}
\end{equation}
 At large $k_y$ when the energy of the surface branch becomes much larger  than $\Delta$, the surface curve  coincides with the one obtained without any strain.
\par
Thus, the surface branch starts at the top of the heavy-hole band and at large $k_y$ approaches the light-hole band, see Fig.~\ref{StrainCase}(a) and similar picture in the Supplementary Material of Ref.~\onlinecite{Brune}. 

\begin{figure}[!ht]
\vspace{0pt}\includegraphics[height=14em]{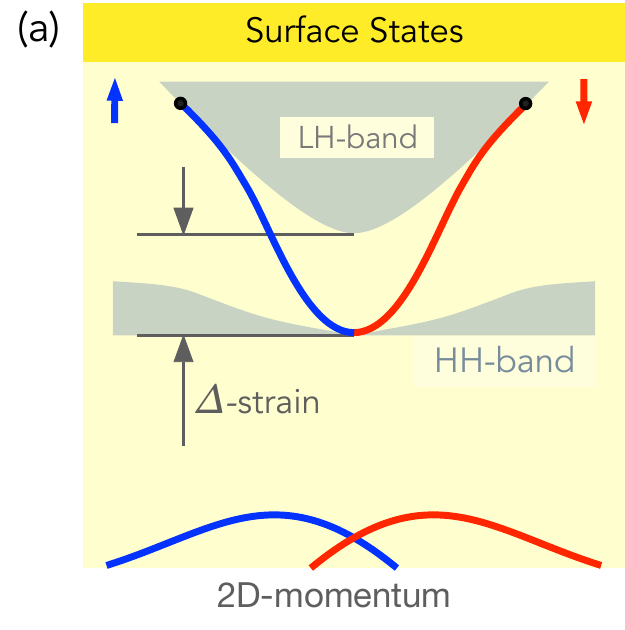}
\vspace{0pt}\includegraphics[height=14em]{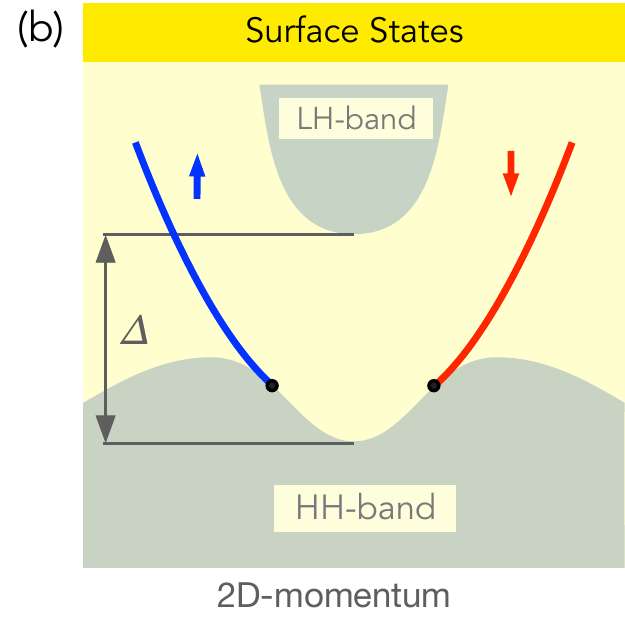}
\caption{\label{StrainCase}
 Effect of a tensile in-plane strain on the surface states studied in two models. 
 (a) Kane model with heavy-hole flat band: 
 The strain splits and intermixes the bulk light-hole and heavy hole bands, opening
 a global gap of magnitude $\Delta/4$ and inducing a dispersion in the heavy-hole band.
 The DK states depart from the heavy-hole band at zero momentum and
 merge with the light-hole band at large momenta. 
 (b) Luttinger model (finite heavy-hole mass but no conduction bands):
 Qualitatively, the same result as in (a), except that the DK states
 depart from the heavy-hole band at finite momenta.
}
\end{figure}

The results of the k-P -model agree very well with the results obtained from more sophisticated methods, like the ab-initio calculations, for Hg-terminated HgTe (001) surface. For Te-terminated surface it is not necessarily the case, see Refs.~(\onlinecite{Rauch,Wu}).

\section{Surface states within Luttinger model with strain}
As we have demonstrated, the surface states that lie within the gap opened by strain between light and heavy-hole bands are of special importance.  These states can be studied in more detail within the Luttinger model which describes the valence $\Gamma_8$ light and heavy-hole bands. This model allows exact analytical results for the spectrum and spinors of the surface states  for an arbitrary ratio between the effective masses of light and heavy holes, and represents a more accurate description compared to the model we used above. In Ref.~\onlinecite{Dyak} the solution for those states was found in the  absence of strain. 
\par
The Luttinger Hamiltonian for HgTe which describes the group of states $|lh -1/2, hh +3/2> $ has the form
\begin{eqnarray}
\hat{H}_L&=&
\begin{bmatrix}
k^2(\gamma-\tilde{\gamma}) -\Delta/4 &\,\,\,\,\,  \sqrt{3}\tilde{\gamma}k_+^2 + \sqrt{3}\Delta/4 \\
\sqrt{3}\tilde{\gamma}k_-^2 + \sqrt{3}\Delta/4  & \,\,\,\,\, k^2(\gamma+\tilde{\gamma}) +\Delta/4
\end{bmatrix}
\label{Lutt}
\end{eqnarray}
The energy is counted from the degeneracy point of light-hole  and heavy-hole bands in the HgTe region in the absence of strain  ($\epsilon_{v1}=0$), $\Delta >0$ is the gap value due to strain, $\hat{k}_{\pm}=\hat{k}_x \pm ik_y$, $\hat{k}_x=-id/dx$, $\hat{k}^2=k_y^2+\hat{k}_x^2 $, and wave vector within the boundary is again directed along the y-axis, $\gamma, \tilde{\gamma}$ are the Luttinger constants which determine the bulk spectra of light and heavy holes in the absence of strain: $\epsilon_{l,h}=k^2(\gamma \pm 2 \tilde{\gamma})$. We assume that $\gamma, \tilde{\gamma}>0$ and $(\gamma-2 \tilde{\gamma})<0, (\gamma+2 \tilde{\gamma}) >0$ (gapless material). As before, the Hamiltonian describing the time-reversed solutions is obtained from Eq.~(\ref{Lutt}) by replacing $k_y$ by $-k_y$.
\par
We consider the HgTe sample which occupies the $x>0$ space and $x=0$ is the boundary with the vacuum or wide gap direct material (infinite barrier). 
We then seek for the surface states located within the gap created by strain.  The  x-components of the  wave vectors corresponding to the light and heavy holes for a given  energy $\epsilon$ of the surface state are given by $k_{l,x}=i\kappa_l$, $k_{h,x}=i\kappa_h$, $\kappa_{l,h} >0$, and the state exponentially decays towards the bulk of HgTe region. The expressions for $\kappa_{l,h}$ are 
\begin{equation}
k_y^2-\kappa^2_{l,h}=\frac{\gamma \epsilon +\tilde{\gamma}\Delta \mp \sqrt{(\gamma \Delta/2+2 \tilde{\gamma}\epsilon)^2+ 3\tilde{\gamma}\Delta k_y^2 (4\tilde{\gamma}^2-\gamma^2)}}{(\gamma^2-4\tilde{\gamma}^2)}
\label{kappa}
\end{equation}
Here  $-$  corresponds to $\kappa_l$ and $+$  to $\kappa_h$. The corresponding two-component spinor wave functions look like
\begin{eqnarray}
\Psi_{l,h}(x)&\propto &e^{ik_yy}e^{-\kappa_{l,h}x} 
\left( \begin{array}{cc}
\sqrt{3}\Delta/4 -\sqrt{3}\tilde{\gamma}(k_y+\kappa_{l,h})^2 & \\
\epsilon +\Delta/4 -(\gamma-\tilde{\gamma}) (k_y^2-\kappa_{l,h}^2) 
\end{array}
\right) \nonumber \\
\label{Psi}
\end{eqnarray}
\subsection{Boundary condition and solution}

The wave function of the surface state is the superposition of the functions Eq.~(\ref{Psi}): $\Psi_s(x)=a\Psi_{l}(x)+b\Psi_{h}(x)$. For an infinite barrier we use zero boundary conditions at $x=0$:  $\Psi_s(x=0)=0$, obtaining the following equation for the energy $\epsilon$ of the surface state:
\begin{eqnarray}
[\epsilon +\Delta/4 -(\gamma-\tilde{\gamma}) (k_y^2-\kappa_{h}^2)] [\Delta/4 -\tilde{\gamma}(k_y+\kappa_{l})^2 ]
 \nonumber \\
=[\epsilon +\Delta/4 -(\gamma-\tilde{\gamma}) (k_y^2-\kappa_{l}^2) ][\Delta/4 -\tilde{\gamma}(k_y+\kappa_{h})^2 ]
\nonumber \\
\label{boundary}
\end{eqnarray}
Positive real values of $\kappa_{l,h}$ found from Eq.~(\ref{kappa}) should be used while solving Eq.~(\ref{boundary}).
\par
The exact solution for the surface state dispersion reads:
\begin{equation}
\epsilon=-\left (\frac{\gamma}{2\tilde{\gamma}}\right )\frac{\Delta}{2}+k_y^2 (\gamma+2\tilde{\gamma})\left [1-\frac{(1+\sqrt{3\beta})^2}{4}\right], 
\label{DKh1}
\end{equation}
where $\beta=(2\tilde{\gamma}-\gamma)/(\gamma+2\tilde{\gamma})>0$ is the ratio of the bulk effective masses of the light and heavy holes. The solution Eq.~(\ref{DKh1}) (we call it further as DK 1 branch) coincides at $\Delta=0$ with the solution obtained in Ref.~\onlinecite{Dyak}. It is interesting that the spectrum in the presence of the strain differs from the one at $\Delta=0$ only by the constant shift proportional to the strain value. Using this dispersion we can find from Eq.~(\ref{kappa}) the values of $\kappa_{l,h}$:
 \begin{equation}
\kappa_{l,h}=\frac{(1+\beta)\sqrt{3}}{4\sqrt{\beta}}k_y \mp \frac{k_y}{\sqrt{3\beta}}\sqrt{\left[1-\frac{(1+\sqrt{3\beta})^2}{4}\right]^2 +\frac{3\beta \Delta}{4\tilde{\gamma}k_y^2}}
\label{kappa1}
\end{equation}
This solution is valid at $k_y>0$. Moreover, from the condition that $\kappa_l >0$ we get the critical value of $k_y^{\star}$, thus the surface state actually exists at $|k_y|>k_y^{\star}$, see  Fig.~\ref{StrainCase}(b). 
\begin{equation}
k_y^{\star}=\sqrt{\frac{\Delta}{\tilde{\gamma}[2+\sqrt{\frac{3}{\beta}}(1-\beta)]}}
\label{star}
\end{equation}
It is easy to see that at this critical $k_y$ the surface state curve merges with the bulk heavy-hole band (corresponding to $k_x=0$). 
In accordance with the results obtained above, in the limit $\beta=0$ (infinite mass of the heavy holes) the critical value $k_y^{\star} \to 0$. Moreover, since $k_y^{\star} $ must be  real, we obtain the following condition: $\beta< 3$. It coincides with the one obtained in Ref. \onlinecite{Dyak} for the case $\Delta=0$. As we know from Ref. \onlinecite{Dyak}, in the case 
$\Delta=0$ there are two  solutions for the surface states. This is true also in the presence  of the strain. For the case of $\Delta  \geq 0$  the second solution (DK 2 branch) exists only at $\beta>1/3$. Since this case does not correspond to real HgTe or $\alpha$-Sn materials, we do not consider it here. However, in the case of $\Delta<0$, which corresponds to the compressive in-plane strain, the situation is qualitatively different. This latter Dirac semimetal case is considered in section \ref{semimetal}.

\subsection{Spin  structure of the surface state}
Let us describe the spin structure of the solution obtained in this section. We have mentioned already that even without strain the  DK surface state \cite{Dyak} has all the features of a topological surface state for a 3D topological insulator:
 momentum-spin locking and absence of back scattering.  Indeed, from Eqs.~(\ref{Psi},\ref{DKh1}) at $\Delta=0$ we find the following  wave function at $k_z=0, k_y>0$
\begin{eqnarray}
\Psi_{\downarrow}(x)= (e^{-\kappa_{l}x}-e^{-\kappa_{h}x} ) \cdot \frac{1}{2\sqrt{1+\beta}}
 \left( 
\begin{array}{cc}
   \sqrt{3}+\sqrt{\beta}    & \\
\sqrt{3\beta}-1 
\end{array}
\right) \nonumber \\
\label{chi}
\end{eqnarray}
Let us calculate the average value of the different components of spin using this function. 
While doing so we should remember that the upper coefficient of the spinor Eq. (\ref{chi}) corresponds to $-1/2$ projection over the z-axis and the lower coefficient to the $+3/2$ projection. Coefficients corresponding to $+1/2, -3/2$ projections are equal to zero. Then using $4\times 4$ matrices $\hat{J}_x,\hat{J}_y,\hat{J}_z$ corresponding to the angular momentum 3/2, we obtain $<J_x>=0, <J_y>=0$, and 
\begin{eqnarray}
<J_z>&=&\frac{1}{4(1+\beta)}\left [\frac{3}{2}(\sqrt{3\beta}-1 )^2 -\frac{1}{2}(\sqrt{3}+\sqrt{\beta} )^2\right] 
\nonumber \\
&=&\frac{\beta-\sqrt{3\beta}}{(1+\beta)}
\label{spin0}
\end{eqnarray}
Thus, the average spin is parallel to the boundary and directed perpendicular to the electron momentum. 
For the opposite momentum direction within the plane one obtains the average spin which is the  opposite sign of the value given by Eq.~(\ref{spin0}); therefore,  one has the momentum-spin locking for this surface state.   
\par
Finally, we present here the result obtained within the Luttinger model for the average spin of the surface electron in the presence of strain, $\Delta >0$. At the conditions $\sqrt{\beta}\ll 1$ and $\Delta/(\tilde{\gamma}k_y^2) <\sqrt{3/\beta}$, we obtain
\begin{equation}
<J_z>=-\sqrt{3\beta}\frac{1}{1- \Delta/(9\tilde{\gamma}k_y^2)}
\label{spin-str}
\end{equation}
As before,  $<J_x>=0, <J_y>=0$, i.e. spin is parallel to the boundary and perpendicular to the momentum direction. 
(At $\Delta=0$ and $\beta \ll 1$ the result  Eq.~(\ref{spin-str}) coincides with the one given by Eq.~(\ref{spin0})).
It is interesting that in contrast to the case without strain, the average spin now increases with the lowering of energy, getting near the $k_y^{\star}$, Eq.~(\ref{star}), the maximum value  which is of order of unity even for $\beta\ll 1$.
\par
We consider the case of isotropic Luttinger model; therefore, the spin-momentum locking is an exact characteristic of the surface states, that  can be probed experimentally.  By adding higher order terms one can create some warping. (In practice,  it can be done by growing film of topological material on the substrate material with the proper symmetry). In the particular case of the "hexagonal warping", Ref.~\onlinecite{Fu}, the component of the spin normal to the plane appears;  however,  the component of the spin parallel to the plane still strongly correlates with the momentum direction. Therefore, in a sense, complete breaking of the locking does not happen.

\section{Surface states in the Dirac semimetal regime (Luttinger model)}
\label{semimetal}

\begin{figure}[!ht]
\vspace{0pt}\includegraphics[width=0.6\columnwidth]{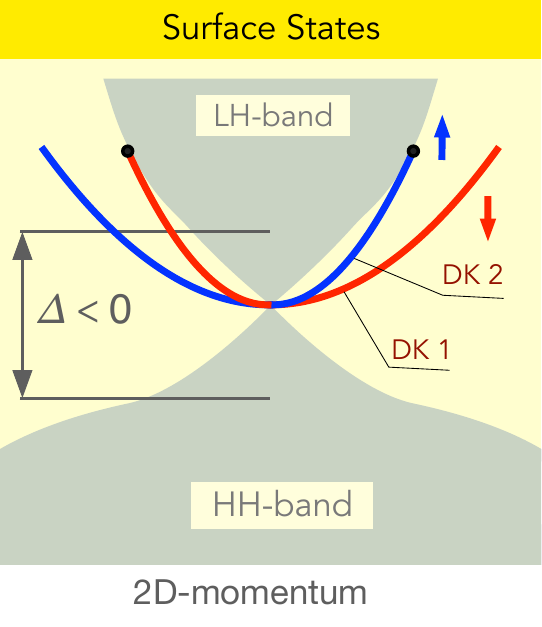}
\caption{\label{semimetDelta}
Case of compressive in-plane strain ($\Delta <0$). Two parabolic surface-state branches (DK 1 and DK 2) 
arise around the conic point of the bulk projected states located at the energy $\epsilon_D=(\gamma |\Delta|/4\tilde{\gamma})$.
The DK 2 branch flows into the bulk spectrum at the value of in-plane momenta given by Eq.~(\ref{tilde_k}).
At $\Delta =0$, the DK 2 branch arises only for $\beta>1/3$, see Ref.~\onlinecite{Dyak}.
}
\end{figure}

So far we have considered the case of tensile in-plane strain ($\Delta >0$), which opens the gap between the light and heavy-hole bands, creating the topological insulator in the material like HgTe.  In the opposite case, $\Delta <0$, one has the situation of the Dirac semimetal where the light-hole band and the heavy-hole band overlap, forming two Dirac points. The bulk energy  dispersion near those points is a linear function of the in-plane  wave vector, see Fig.~(\ref{semimetDelta}).  The case of the Dirac semimetal has received much less attention in the literature, especially its analytical treatment. We consider here the problem of surface states in the Dirac semimetal regime within the Luttinger model. Similar to the $\Delta>0$ case, it is possible to find an exact analytical solution for the spectrum and wave functions of the surface states within the Luttinger model (i.e. for energies smaller than the $\Gamma_8-\Gamma_6$  gap).  As far as we know, it is the first time when a comprehensive analytical solution for this problem is found. 
Note, that we do not study here the surface states located within the  $\Gamma_8-\Gamma_6$ gap. It is clear, however, that similar to the previous cases considered in this paper, those states  lie on top of the continuum
of the heavy-hole states and are strongly  hybridized  with them. Therefore, 
 there are no linear in k-vector surface states going from s-band to the light-hole band across the $\Gamma_8-\Gamma_6$ gap,  i.e. the popular picture is again incorrect. 
\par
We start with the bulk spectrum near the Dirac point  at $\Delta <0$.  Again the $x$-axis is perpendicular to the boundary, and the wave vector along the boundary is equal to $k_y$. From the expression for the dispersions of the light and heavy holes
\begin{equation}
\epsilon_{l,h}=\gamma (k_x^2+k_y^2) \pm 2\tilde{\gamma}\sqrt{ \left (\frac{\Delta}{4\tilde{\gamma}}+k_x^2\right )^2 +k_y^2 \left (k_y^2 +2k_x^2 -\frac{\Delta}{4\tilde{\gamma}}\right ) }
\label{bulk1}
\end{equation}
one can easily see that at $k_x^2=|\Delta |/4\tilde{\gamma}$ (Dirac point) the bulk spectrum  is linear
\begin{equation}
\epsilon_b (k_y)=\frac{\gamma |\Delta |}{4\tilde{\gamma}} \pm k_y \sqrt{3 \tilde{\gamma} |\Delta |}; \,\,\, |k_y| \ll 
\sqrt{|\Delta |/ \tilde{\gamma} }.
\label{Dirac1}
\end{equation}
We consider now the surface states. We will show that the two branches of parabolic surface states start from the bulk Dirac  point. 
It is easy to realize that we can use the same Eq.~(\ref{boundary}) for the dispersion of the surface states. The only difference now  is that the strain is negative: $\Delta<0$. Therefore, the solutions Eqs.~(\ref{DKh1},\ref{kappa1}) are still valid.
We see that the first (DK 1) branch, Eq.~(\ref{DKh1}), starts at the energy of the bulk Dirac point: $\gamma |\Delta |/4\tilde{\gamma}$, see Fig.~(\ref{semimetDelta}), and has the same mass in the absence of strain. The expressions for the $\kappa_{l,h}$ are still given by Eq.~(\ref{kappa1}), the only difference from the case $\Delta >0$ is that now quantities  $\kappa_{l,h}$ become complex at small enough values of $k_y$. Consider the positive $k_y$,  the character of the surface wave function is different depending on  whether $k_y$ is smaller or bigger than $k_{y1}$, where 
\begin{equation}
k_{y1} = 2\sqrt{\frac{\beta |\Delta |}{ \tilde{\gamma}}}\cdot \frac{1}{|1-\sqrt{3\beta}|( \sqrt{3}+\sqrt{\beta} ) }.
\end{equation}
If $k_y<k_{y1}$, then $\kappa_l=\kappa_h^{\star}=a-ib$, and the surface function has the form $\Psi_s(x)\propto \sin(bx)\cdot \exp(-ax)$ with $a,b$ to be found from Eq.~(\ref{kappa1}).  In the case $k_y>k_{y1}$ both $\kappa_{l,h}$ are real and positive,  and the surface wave function has the form $\Psi_s(x)\propto (e^{-\kappa_{l}x}-e^{-\kappa_{h}x} ) $, with $\kappa_{l,h}$ given by Eq.~(\ref{kappa1}) at $\beta <3$. 
\par
The interesting feature of the problem is that  there is a great difference in the number of DK surface branches which exist at positive or negative $\Delta$. In the case of $\Delta >0$, at $\beta <1/3 $ (which is the case of HgTe and $\alpha$-Sn materials) there is only one branch  DK 1 of surface states. We will show now that 
in the case of $\Delta <0$, considered here, there are always {\it  two} DK surface branches (even for $\beta <1/3$).  Thus, in the case of Dirac semimetal, one always has two DK surface branches, which for the unstrained sample were found in the original work Ref.~(\onlinecite{Dyak}).  We should mention that the authors of Ref.~(\onlinecite{Kibis})  found only one DK branch in the regime of a Dirac semimetal. Moreover, these two branches continue until the $k_y=0 $ point, in strong contrast to the conclusion made in Ref.~(\onlinecite{Kibis}). 
\par 
Let us discuss the second surface branch, which exists at  $\Delta <0$. It is the solution of Eq.~(\ref{boundary}) where the change $k_y \rightarrow -k_y$ is made, i.e. for positive $k_y$ it is obtained from the second  block of basis states and has an opposite spin compared to the first (DK 1) solution. The exact solution for the dispersion of the  second surface branch (DK 2) reads:
\begin{equation}
\epsilon=\left (\frac{\gamma}{2\tilde{\gamma}}\right )\frac{|\Delta|}{2}+k_y^2 (\gamma+2\tilde{\gamma})\left [1-\frac{(1-\sqrt{3\beta})^2}{4}\right].
\label{DKh2}
\end{equation}
The corresponding values of $\kappa_{l,h}$ are:
 \begin{equation}
\kappa_{l,h}=\frac{(1+\beta)\sqrt{3}}{4\sqrt{\beta}}k_y \mp \frac{k_y}{\sqrt{3\beta}}\sqrt{\left[1-\frac{(1-\sqrt{3\beta})^2}{4}\right]^2 +\frac{3\beta \Delta}{4\tilde{\gamma}k_y^2}}
\label{kappa2}
\end{equation}
This solution is valid at $k_y>0$. For small  enough $k_y$ the quantities  $\kappa_{l,h}$ are complex  again, and one can have a localized state even at $\beta <1/3$, in strong contrast to the cases $\Delta \geq 0$. At large enough $k_y$, however, when both   $\kappa_{l,h}$ are real, from the condition that   $\kappa_l >0$ we get the critical value $k_y=\tilde{k}_y$,  after which the second surface branch ceases to exist (it merges with the bulk band), see  Fig.~(\ref{semimetDelta}). 
The expression for $\tilde{k}_y$ reads
 \begin{equation}
\tilde{\gamma}\tilde{k}_y^2= \frac{\sqrt{\beta} |\Delta |}{(1-\sqrt{3\beta}) ( \sqrt{3}+\sqrt{\beta} )}; \,\,\, \beta <1/3. 
\label{tilde_k}
\end{equation}

Thus we see that the dispersions of the surface states found by us here are very much different from the ones usually used, see,  for example, Refs.~(\onlinecite{Ding,Ding1}). The authors of   Refs.~(\onlinecite{Ding,Ding1}) came to the conclusions that the effects they observe  are due to the Dirac semimetal surface states of $\alpha$-Sn layer rather than due to the bulk states of this material. 
One of the effects is the switching of a magnet by spin-orbit torque from  a Dirac semimetal by a charge current at room temperature, without an external magnetic field. This shows that the  topological Dirac semimetal    $\alpha$-Sn is a promising material for the spintronic applications. In that respect, the results obtained by us are potentially important for the correct interpretation of the current and future experiments. 
 
\section{Conclusions} 
We have reconsidered the physical origin of surface states in inverted-band semiconductors, such as HgTe and $\alpha$-Sn. 

By  considering analytically two simple and exactly solvable models we have clarified several important issues concerning the nature of the states, among them the role of band structure and the role of strain. In particular, we have shown that neglecting the coupling to the heavy-holes band  leads to the incorrect physical picture of the surface states obtained in many recent publications. Such a procedure results in the simple picture of linear in k-vector Dirac surface states crossing the whole gap between the conduction and light-hole bands, see Fig.~\ref{bands}(b). This latter picture does not reflect the true nature of the surface states.
We have also shown that in a topological insulator regime an applied  strain is a smooth “deformation” to the surface states that does not lead to any drastic change of the physical properties in these materials, in contrast to what has recently been published in the literature. In the Dirac semimetal regime, however,  the physics of surface states  which arise around the conic point of the bulk projected states crucially depends on the applied strain. 

We have reconciled different analytical solutions found in the 1980s with the results obtained recently by many groups 
using numerical simulations. In particular, we have demonstrated that the DK surface state \cite{Dyak} predicted for HgTe for
the first time within the Luttinger model is the result of strong hybridization of the Dirac state with the heavy-hole states.
We have also shown that these states \cite{Dyak} possess all the essential features of the topological states in the
presence of strain that induces a gap between the light and heavy-hole bands. As such, they should play the
most important role in the experiments that probe the topological properties of the sample when the Fermi level
is located within this strain-induced gap. In contrast, the surface states located
within the  $\Gamma_8-\Gamma_6$  gap lie on top of the continuum
of the heavy-hole states   and strong hybridization with them  makes those surface states marginal
in observable phenomena.  They still can be detected with the ARPES experiments at small 2D momenta along the interface where the overlap of these states with heavy holes is weak, see Refs.~(\onlinecite{Rogalev,Rauch1}). 
\par 
The results obtained by us can be used  for the correct interpretation of the current and future experiments with materials such as $\alpha$-Sn and HgTe.

This work is supported by the Air Force Office of Scientific Research (FA9550-AFOSR-20RYCOR039). 
 This research was performed while A. Khaetskii held an NRC Research Associateship award at the Sensors Directorate, Air Force Research Laboratory. This work was  performed in part at Aspen Center for Physics, which is supported by National Science Foundation grant PHY-1607611.  A. Khaetskii acknowleges the hospitality of the ACP.

\end{document}